\documentclass[12pt]{article}
\usepackage{amsxtra,amssymb,amsthm,amsmath,latexsym}
\theoremstyle{plain}

\def\nd{\noindent}

\def\ve{{\varepsilon}}

\def\R{{\mathbb R}}

\def\calL{{\mathcal L}}
\def\calU{{\mathcal U}}

\def\tildeC{{\widetilde C}}
\def\tildeD{{\widetilde D}}
\def\diam{{\hbox{\,diam\,}}}

\def\barq{{\overline{q}}}
\def\barv{{\overline{v}}}
\def\bee{\begin{equation*}}
\def\eee{\end{equation*}}
\def\be{\begin{equation}}
\def\ee{\end{equation}}
\DeclareMathOperator{\mindist}{min\ dist}

\begin{document}
\title{ Wave scattering by small impedance particles in a medium 
}
\author{A.G. Ramm\\
 Mathematics Department, Kansas State University, \\
 Manhattan, KS 66506-2602, USA\\
ramm@math.ksu.edu}
\date{}
\maketitle\thispagestyle{empty}
\begin{abstract}
\footnote{MSC: 35J05, 35P25, 73D25, 81U10, 82D20}
\footnote{PACS: 0304K, 43.20.tg, 62.30.td}
\footnote{key words: wave scattering, small particles, many-body 
problems, metamaterials}

The theory of acoustic wave scattering by many small bodies is developed
for bodies with impedance boundary condition. It is shown that if one
embeds many  small particles in a bounded domain, filled with a known
material, then one can create a new material with the properties 
very
different from the properties of the original material. 
Moreover, these very different properties occur
although the total volume of the embedded small particles is negligible
compared with the volume of the original material.
\end{abstract} 

\section{Introduction}\label{S:1}
In \cite{R524} a theory of wave scattering by many small acoustically
soft particles, embedded in a bounded domain $D$, filled with
a material with known refraction coefficient, a medium,  is 
developed. "Acoustically soft"
means that the Dirichlet condition holds on the boundary of the small
particles. Using the general methodology, developed in \cite{R524}, we
study here the wave scattering on impedance particles, derive a linear
algebraic system for quantities which
 yield the scattered field if the
number $M$ of the embedded particles is of order 10, and a linear
integral equation for the self-consistent (effective) field in the medium,
consisting of the medium in which many ($M\to\infty$) small 
particles are 
embedded, if suitable physical assumptions are made, which 
include
the following assumptions:
\be\label{e1}
  a<<\lambda, \qquad a<< d, \ee
where $a$ is the characteristic size of a small particle, $\lambda$ is
the wavelength in the medium, $d$ is the smallest distance between any
two distinct particles. We prove that the embedded particles create a new
material whose refraction coefficient (in the limit $M\to\infty$) can be
an arbitrary desired function, although the total volume of the embedded
particles tends to zero as $M\to\infty$. Thus, our theory may lead to a
new technology in creating materials with desired properties by embedding
into original material many small particles with the number of particles
per unit volume around any point $x$ as well as their impedances 
calculated so that the resulting new material would have a desired
refraction coefficient $n(x)$.
The embedding of the small particles can be done 
using nanotechnology.

For the theory of wave scattering by small bodies, originated by Rayleigh
in 1871, we refer to \cite{LL}, \cite{DK} and \cite{R476}.

\section{Statement of the problem and its solution.}\label{S:2}
Let $D\subset\R^3$ be a bounded domain filled with a known material with
refraction coefficient $n_0(x)$, $x\in\R^3$. The scattering problem
consists of finding the solution to the equation
\be\label{e2}
  \calL u:=[\nabla^2+k^2n_0(x)]u=0\hbox{\ in\ } 
\R^3, \ee
\be\label{e3} u=u_0+A_0(\beta,\alpha) \frac{e^{ikr}}{r} 
  +o\left(\frac{1}{r}\right),
  \quad r:=|x|\to\infty, \quad \beta:=\frac{x}{r}, \ee
\be\label{e4}
  u_0=e^{ik\alpha\cdot x}, \ee
$\alpha\in S^2$ is the unit vector in the direction of the incident wave,
$\beta$ is the unit vector in the direction of the scattered wave, $k>0$
is fixed throughout the paper, $A_0(\beta,\alpha)$ is the scattering
amplitude, its dependence on $k$ is not shown since $k$ is fixed. 

We
assume that 
\be\label{e5}
  n_0(x)=1\hbox{\ in\ } D':=\R^3\backslash D, \ee
and write the Schr\"odinger equation, equivalent to \eqref{e2}:
\be\label{e6}\calL u=[\nabla^2+k^2-q_0(x)]u=0\hbox{\ in\ } \R^3,
 \qquad q_0(x):=k^2(1-n_0(x))=0\hbox{\ in\ } D'. \ee
We assume that $q_0$ is a bounded piecewise-continuous in $\R^3$
function. 
Problem \eqref{e3}, \eqref{e4}, \eqref{e6} has a unique solution for any
square-integrable real-valued $q_0\in L^2(D)$, or for complex-valued $q$
with $Im\,q\leq 0$. We sketch a proof in the Appendix. 

Denote by $u_0$
the solution to the scattering problem \eqref{e2}-\eqref{e3}, and by $G$
the corresponding Green's function, which solves the following problem:
\be\label{e7}
  \calL G=-\delta(x-y)\hbox{\ in\ } \R^3, \ee
\be\label{e8}
  \frac{\partial G}{\partial |x|} -ikG=o\left(\frac{1}{|x|}\right), 
  \qquad |x|\to\infty, \ee
where $\delta(x-y)$ in \eqref{e7} is the delta function.
Consider now $M$ small particles $D_m$, $1\leq m\leq M$, embedded in $D$,
and the scattering problem:
\be\label{e9}
   \calL\calU=0\hbox{\ in\ } \R^3\backslash U^M_{m=1} D_m, \ee
\be\label{e10}
  \calU=\calU_0+V, 
  \qquad \frac{\partial V}{\partial |x|}-ik
V=o\left(\frac{1}{|x|}\right),
  \qquad |x|\to\infty, \ee
 \be\label{e11}
  \frac{\partial\calU}{\partial|x|}=h_m\calU \hbox{\ on\ } S_m, 
  \quad 1\leq m\leq M,\ee
where $N$ is the unit exterior normal to $S_m$, $S_m$ is the boundary of
$D_m$. We assume that $S_m$ is Lipschitz uniformly with respect to $m$.
In the impedance boundary condition \eqref{e11} the parameter $h_m$ is
a constant, possibly complex-valued, such that problem 
\eqref{e9}-\eqref{e11} has a unique solution. For example, this is the case if
$Im\, h_m\leq 0$.

Let us look for this solution of the form
\be\label{e12}
  \calU=\calU_0+\sum^M_{m=1} \int_{S_m} G(x,t)\sigma_m(t)dt, \ee
where $\sigma_m$ should be chosen so that the boundary condition
\eqref{e11} is satisfied. For arbitrary $\sigma_m\in L^2(S_m)$ the
function \eqref{e12} solves equation \eqref{e9} and satisfies conditions
\eqref{e10}.
Therefore, if $\sigma_m$ are found so that conditions \eqref{e11} are
satisfied, then \eqref{e12} solves problem \eqref{e9}-\eqref{e11}. So far
we did not use the assumption that $D_m$ are small. Let us assume
\eqref{e1}, where
\be\label{e13}
  a=\frac12 \max_{1\leq m\leq M} \diam D_m, 
  \qquad d=\underset{1\leq m,j\leq M,\, j\not= m}{\mindist}  (D_m,D_j),\ee
and $\lambda=\frac{2\pi}{k\sqrt{n_0}}$ is the wavelength in the medium
with the refraction coefficient $n_0(x)$. We assume that $n_0(x)$ is
practically constant on the distances of the order $\lambda$.

Let us denote
\be\label{e14}
  Q_m:=\int_{S_m} \sigma_m(t)dt, \qquad 1\leq m\leq M,\ee
and rewrite \eqref{e12} as:
\be\label{e15}
  \calU=\calU_0+\sum^M_{m=1} G(x,x_m)Q_m
  +\sum^M_{m=1}\int_{S_m}[G(x,t)-G(x,x_m)] \sigma_m(t)dt, \ee 
where $x_m\in D_m$ can be chosen arbitrary because $D_m$ is small. One
may take $x_m$ to be the gravity center of $D_m$. The gravity center
$x_m$ may lie outside $D_m$ if $D_m$ is not convex, but $x_m$ belongs to
the convex hull of $D_m$. We assume that $D_m$ are convex, but this
assumption is not essential. If $Q_m$ does not vanish we expect that
\be\label{e16}
  |G(x,x_m)Q_m|>> \bigg| \int_{S_m}[G(x,t)-G(x,x_m)]\sigma_m(t)dt\bigg|,
  \quad \hbox{\,\,\, if\,\,\,}\quad |x-x_m|>d>>a.\ee
We assume that \eqref{e16} holds for $|x-x_m|>>a$ if $Q_m\not=0$ because
\be\label{e17}
  G(x,t)-G(x,x_m)=\nabla_y\ G(x,y)\big|_{y=x_m+\tau(t-x_m)}\cdot (t-x_m),
  \qquad 0<\tau<1,\ee
so that, for a fixed $k>0$, one has
\be\label{e18}
  |G(x,t)-G(x,x_m)|=O\left(\frac{a}{|x-x_m|}\right) << 1. \ee
Therefore the integral in \eqref{e16} is $O(ka|G(x,x_m)Q_m|)$
if $Q_m\not= 0$, $|x-x_m|>>a$, and $ka<<1$.

In this approximation we have
\be\label{e19}
  \calU(x)= \calU_0(x) +\sum^M_{m=1} G(x,x_m) Q_m,
  \qquad |x-x_m|>d>>a\quad \forall m, \ee
with an error of order $O(ka+\frac a d)$.

Since the potential $q_0(x)$ is known, one may consider 
the functions $\calU_0(x)$ and $G(x,y)$ 
known.
Therefore the scattering problem \eqref{e9}-\eqref{e11} is solved if one
finds $Q_m$, $1\leq m\leq M$. 

Let us derive the equations for finding
$Q_m$, 
$1\leq m\leq M$, using the boundary condition \eqref{e11}. Denote the
effective field, acting on the $j\hbox{-th}$ particle by $\calU_e$, where
\be\label{e20}
  \calU_e(x):=\left\{
  \begin{array}{ll}
  \calU_0(x)+\sum^M_{m\not= j} G(x,x_m)Q_m, & |x-x_j|\sim a,\\
  \calU_0(x)+\sum^M_{m=1} G(x,x_m)Q_m, & |x-x_m|>d>>a\quad\forall m.
  \end{array}\right.\ee
Condition \eqref{e11} yields:
\be\label{e21}
 \left(\frac{\partial}{\partial N}-h_j\right) \int_{S_j} G(x,t)
  \sigma_j(t)dt
  =-\left(\frac{\partial}{\partial N}-h_j\right) \calU_e (x), \ee
where $x$ is a point on $S_j$ and the normal derivative on $S_j$ is taken 
from the exterior to $D_j$ domain in \eqref{e21} and below.
Using formulas  (A.8) and (A.15) from the Appendix one may 
replace the function $G(x,t)$ in the
integral in equation
\eqref{e21} by the function $g_0(x,t)=\frac 1{4\pi |x-t|}$ with the error 
of order $O(ka)$.
Using the known formula for the normal derivative of a single-layer 
potential (see, e.g.,  \cite[p.5]{R476} ):
\be\label{e22}
  \frac{\partial}{\partial N} \int_{S_j} g_0(x,t)\sigma_j(t)dt
  =\frac{A_j \sigma_j-\sigma_j}{2}, \quad
  \quad A_j \sigma_j
  := 2\int_{S_j} \frac{\partial g_0(s,t)}{\partial N_s} \sigma_j(t)dt,
\ee
one rewrites \eqref{e21} as
\be\label{e23}
  \sigma_j(s)=A_j\sigma_j-2h_j T_j \sigma_j
  +2\frac{\partial\calU_e(s)}{\partial N}-2h_j\calU_e(s),\quad s\in
S_j,\ee
where
\be\label{e24}
  T_j\sigma_j:=\int_{S_j} g_0(s,t)\sigma_j(t)dt. \ee
It is known (see \cite[p.96]{R476}) that
\be\label{e25}
  \int_{S_j} ds\ A_j\sigma=-\int_{S_j} \sigma\,dt. \ee
Integrate \eqref{e23} over $S_j$, use \eqref{e14} and \eqref{e25},
and take into account that
\be\label{e26}
  \int_{S_j} \frac{\partial \calU_e}{\partial N} \,ds
  =\int_{D_j} \Delta \calU_e dx\approx V_j \Delta\calU_e(x_j),
  \quad V_j=|D_j|, \ee
where we have used the smallness of $D_j$ to replace the integral
approximately by the expression $ V_j \Delta\calU_e(x_j)$ and denoted
by  $|D_j|$ 
 the volume $V_j$ of $D_j$. Thus, the integration of 
\eqref{e23} over $S_j$ yields:
\be\label{e27}
  Q_j=-h_j\int_{S_J}ds\ \int_{S_j}\frac{\sigma_j(t)dt}{4\pi|s-t|}
  -h_j\int_{S_j} \calU_e(x)ds+V_j\Delta\calU_e(x_j). \ee
Note that 
\be\label{e28}
  \int_{S_j} dt \sigma_j(t)\int_{S_j}\frac{ds}{4\pi|s-t|}
  \approx \int_{S_j} dt\sigma_j(t)\frac{1}{|S_j|} 
  \int_{S_j} dt\int_{S_j}\frac{ds}{4\pi|s-t|} =Q_j\frac{J}{4\pi|S_j|},
\ee
where $|S_j|$ is the surface area of $S_j$,
\be\label{e29}
  J:=\int_{S_j}\int_{S_j}\frac{dsdt}{|s-t|},\quad Q_j=\int_{S_j}\sigma_j(t)dt, 
\ee
and we have replaced the function $\int_{S_j}\frac{ds}{4\pi|s-t|}$, 
practically constant at the distances of
order $a$, by its mean value 
$$\frac 1 {|S_j|}\int_{S_j}dt\int_{S_j}\frac{ds}{4\pi|s-t|}$$ 
i.e., by $\frac{J}{4\pi |S_j|}$. Let us estimate the order of smallness of 
various terms in \eqref{e27}. The term 
$\int_{S_j}\calU_e(x)ds\approx \calU_e(x_j) |S_j|=O(a^2)$,
while the term $V_j\Delta\calU_e(x_j)=O(a^3)$, if we assume that
$\calU_e$ and $\Delta\calU_e$ are bounded. Therefore, we neglect the last
term on the right in \eqref{e27}, and obtain
\be\label{e30}
  Q_j=-\frac{h_j|S_j|\calU_e(x_j)}{1+\frac{h_j J}{4\pi|S_j|}}. \ee
In \cite[p.27]{R476}, the following approximate formula is derived for the
electric capacitance $D_j$ of the perfect conductor with the surface
$S_j$:
\be\label{e31}
  C_j\simeq \frac{4\pi|S_j|^2}{J}, \ee
where the conductor is placed in a medium with the dielectric constant 
$\ve_0=1$. Using \eqref{e31}, one may rewrite \eqref{e30} as:
\be\label{e32}
  Q_j=-\frac{C_j}{1+\frac{C_j}{h_j|S_j|}} 
  \calU_e(x_j):=- \tildeC_j\calU_e (x_j),\quad 
\tildeC_j:=\frac{C_j}{1+\frac{C_j}{h_j|S_j|}}. \ee
When $h_j\to\infty$, that is, when the impedance boundary condition 
becomes the Dirichlet condition in the limit $h_j\to\infty$, then  
one obtains from \eqref{e32} the familiar relation 
$Q_j=-C_j\calU_e(x_j)$ for the total charge $Q_j$ on the surface of the
perfect
conductor $D_j$ charged to the potential $-\calU_e(x_j)$. Here
$\calU_e(x_j)$ is defined in \eqref{e20} and on the distances of order 
$a$ the field $\calU_e(x)$
is practically constant.

Substitute \eqref{e32} into \eqref{e20}, multiply by $C_j$,  set $x=x_j$,
and get
\be\label{e33}
   Q_j=-\tildeC_j\calU_0(x_j) -\sum^M_{m\not=j} G(x_j,x_m) \tildeC_j Q_m,
\qquad 1\leq j \leq M.
\ee 
This is a linear algebraic system for finding the unknown $Q_m$, $1\leq
m\leq M$.
The matrix of this system is diagonally dominant if
\be\label{e34}
 \max_{1\leq j\leq M}\sum_{m\not= j} |G(x_j,x_m)| |\tildeC_j|<1. \ee
If condition  \eqref{e34} holds, then system \eqref{e33} can be solved by
iterations:

\be\label{e35}
  Q_j^{(n+1)} =-\tildeC_j \calU_0(x_j) - \sum^M_{m\not= j} G(x_j,x_m)
\tildeC_j
  \ Q^{(n)}_m,\quad Q^{(0)}_j=-\tildeC_j\calU_0(x_j), \ee
and this iterative process converges at the rate of a geometric series.
Therefore, system \eqref{e33} is convenient for solving the scattering 
problem 
\eqref{e9}-\eqref{e11} when the number of small particles is not very
large, $M=O(10^3)$. If this number is very large ($M\sim~ 10^{23}$),
then we study the limiting behavior of $\calU_e(x)$ as $M\to\infty$ and
derive an integral equation for the effective field $\calU_e$ in the
resulting continuous medium. 

We rewrite the second line of \eqref{e20} using formula \eqref{e32}
and obtain the following representation for $\calU_e(x)$:
\be\label{e33a}
 \calU_e(x)=\calU_0(x)-\sum^M_{m=1} G(x,x_m)\tildeC_m\calU_e(x_m). \ee
Assume now that the limiting density of the quantities $\tildeC_m$ exists
in the following sense: if $\tildeD\subset D$ is an arbitrary subdomain
of $D$, then there exists the following limit:
\be\label{e34a}
  \int_{\tildeD}\tildeC(x)dx=\lim_{M\to\infty} \sum_{D_m\subset\tildeD} 
  \tildeC_m. \ee
Under this assumption one can pass to the limit $M\to\infty$ in 
\eqref{e33a}
and get 
\be\label{e35a}
  \calU_e(x)=\calU_0(x)-\int_D G(x,y)\tildeC(y)\calU_e(y)dy. \ee
Applying the operator $\calL$, defined in \eqref{e6}, to \eqref{e35a} and
using \eqref{e7}, one gets
\be\label{e39}
{\cal L} \calU_e-\tildeC(x)\calU_e=0. \ee
This is a Schr\"odinger equation with the potential
\be\label{e40}
  q(x):=q_0(x)+\tildeC(x). \ee
The corresponding scattering amplitude is
\be\label{e41}
  A(\beta,\alpha)=A_0(\beta,\alpha)-\frac{1}{4\pi} \int_D\calU_0(y,-
\beta)
  \tildeC(y)\calU_e(y)dy,  \qquad \calU_e(y)=\calU_e(y,\alpha), \ee
where $\alpha\in S^2$ is the unit vector in the direction of the incident 
plane wave,   the following formula (see \cite[p.25, 
formula(5.1.7)]{R470}) was used:
\be\label{e42}
  G(x,y)=\frac{e^{ik|x|}}{4\pi|x|}\calU_0(y,-\alpha)
  +o\left(\frac{1}{|x|}\right), \qquad |x|\to\infty,\quad
\frac{x}{|x|}=\alpha,\ee
and $\calU_0(x,\alpha)$ is the scattering solution, i.e., the solution to
problem \eqref{e2}-\eqref{e4} (or, which is the same, to problem
\eqref{e3}, \eqref{e4}, \eqref{e6}).

To calculate the scattering amplitude by formula \eqref{e41} one has to
first solve the integral equation \eqref{e35a} for 
$\calU_e$.
The amplitude $A_0(\beta,\alpha)$ in \eqref{e41} is the same as in
formula \eqref{e3}.

If the number of particles is not very large, one may use formula
\eqref{e20} and calculate the scattering amplitude by the formula
\be\label{e43}
  A(\beta,\alpha)=A_0(\beta,\alpha)+\frac{1}{4\pi}\sum^M_{m=1}
 \calU_0(x_m,-\beta)Q_m, \qquad Q_m=Q_m(\alpha),\ee
where the quantities $Q_m$ are found from the linear algebraic system 
\eqref{e33}.

\section{Possible applications to constructing metamaterials.}\label{S:3}

Our idea for creating a material with a desired refraction coefficient is 
simple:  such a material can be obtained by embedding many
small particles in the original material, which fills the domain $D$.
The embedding should create the desired refraction coefficient 
\be\label{e44}
  n(x):=1-k^{-2}q(x), \quad \hbox{\,\,\, where \,\,\,} \qquad 
q(x)=q_0(x)+\tildeC(x),\ee
compare with formula \eqref{e6}. The function $\tildeC(x)$ is fairly
general, so the new material, that we have obtained from the original
one, which has  refraction coefficient $n_0(x)$, is
rather general. 

Let us prove that the total volume of the embedded small
particles is negligible compared with the volume of the original material
in the domain $D$, 
although the effect, produced by these particles
on the refraction coefficient, is large. Consider a unit cube in $D$, 
filled with the
original material. If the distance between two distinct small particles is not
less than $d$, then the number of these particles in this unit cube is not
greater than $\frac{1}{d^3}$, and the total volume of the embedded small
particles in this cube is $O(\frac{a^3}{d^3})$. If
$M\to\infty$ and  $\frac a d\to 0,$  then 
$\frac{a^3}{d^3}\to 0$, so that the limit of the relative volume of the 
embedded small particles per unit volume of the original material in 
$D$ {\it is zero}.

How is it possible that these particles produce large effect on the
refraction coefficient? How is it possible that 
$\tildeC(y)\not\equiv 0$?

The reason is simple. Let us give a simple calculations in order to 
explain 
this reason. Suppose, for simplicity, that the small particles in
a unit volume of $D$ are identical. Then the limit \eqref{e34a} exists
and is not zero, provided that the limit 
\bee \lim_{M\to\infty}\ \frac{a}{d^3}=\tildeC \eee
exists and $\tildeC\not\equiv 0$. Indeed, $O(\frac{1}{d^3})$ is the
order of the number of
small particles per unit volume, and, by equation \eqref{e32},
\be\label{e45}
  \tildeC_m=\frac{C_m}{1+\frac{C_m}{h_m|S_m|}}=O(a), \ee 
because $C_m=O(a)$ and the quantity $C_m{h_m|S_m|}$ can be made $O(1)$
by choosing $h_m$ properly. Note that the dimension $[h_m]=L^{-1}$, where
$L$ stands for length, $[C_m]\sim a$, $[a]=L$, $[S_m]\sim a^2$,
$[S_m]=L^2$, so the 
quantity
$\frac{C_m}{h_m|S_m|}$ is dimensionless.

{\it To summarize:} 

One can have 
$\lim_{\substack{M\to\infty\\d\to 0}} \left(\frac{a}{d}\right)^3=0$ and 
$\lim_{\substack{M\to\infty\\d\to 0}} \frac{a}{d^3}=\tildeC\not\equiv 0$,
provided that $d=O(a^{1/3}\gamma)$, where $\gamma$ is a
bounded 
coefficient whose dimension is $L^{2/3}$, so that 
the two quantities, $d$ and $a^{1/3}\gamma$,
have the same dimension $L$.

Suppose that all the embedded in $D$ small particles have the same shape
and, possibly, different impedances $h_m$. Let $N(x)$ be the density of
the number of these particles per unit volume around point $x$, that is,
for any subdomain $\tildeD\subset D$ the following limit exists:
\be\label{e46}
  \int_{\tildeD}N(x)dx=\lim_{M\to\infty} \sum_{D_m\subset \tildeD}1. \ee
One can also write 
$$N(x)dx=\sum_{D_m\subset dx}1,$$ where $dx$ is a small
element of volume around point $x\in D$, such that $dx$ still contains
many small particles. 
Then
\be\label{e47}
  \tildeC(x)\approx N(x)\tildeC_m=\frac{N(x)C}{1+\frac{C}{h(x)|S|}}. \ee
Here $C$ is the electric capacitance of a perfect conductor with the
shape of a single small particle, $|S|$ is the surface area of this
particle, and $h(x)$ is the boundary impedance of small particles around
point $x$. If $h(x)=h_1(x)+ih_2(x)$, where $h_1$ and $h_2$ are 
arbitrary real-valued functions, then the function $\tildeC(x)$ is also
arbitrary. In formula \eqref{e47} the three function $N(x)\geq 0$, 
$h_1(x),$
and $h_2(x)$ can be chosen to produce a desired function $\tildeC(x)$,
that is, to produce the desired refraction coefficient $n(x)$ by formula
\eqref{e44}.

Let us give the condition on $h_2$ and $q_0(x)$ that are sufficient for
the uniqueness of the solution to the scattering problem \eqref{e9}-
\eqref{e11}. In Appendix some sufficient conditions for this uniqueness
have been established, namely:
\be\label{e48}
  Im\,q(x)\leq 0, \qquad Im\,h=h_2\leq 0. \ee

If $\tildeC(x)=\tildeC_1(x)+i\tildeC_2(x)$, where $\tildeC_1(x)$ and
$\tildeC_2(x)$ are real-valued functions, then solving 
\eqref{e47} for $h(x)$ yields:
\be\label{e49}
  h(x)= \frac{C\tildeC(x)}{|S|[N(x)C-\tildeC(x)]},
  \quad  Im\,h=\frac{C}{|S|} 
 \frac{\tildeC_2 CN(x)}{[N(x) C-\tildeC_1(x)]^2+\tildeC^2_2(x)}. \ee
Thus, if $Im\,h\leq0$ then $Im\,\tildeC_2\leq 0$. Inequality $Im\,q\leq0$
holds if $Im\,q_0(x)+\tildeC_2(x)\leq 0$.
Therefore, if one wants to create a material with the desired refraction 
coefficient $n(x)$,
i.e., with a desired $q(x)=k^2-k^2n(x)$, $Im\,q(x)\leq 0$, then one
starts with an arbitrary $n_0(x)$, $Im\,n_0=0$, i.e., with 
$q_0(x)=k^2-k^2n_0(x)$. Given $q_0(x)$ and $q(x)$, one finds 
$\tildeC(x)=q(x)-q_0(x)$, and then uses formula \eqref{e49} 
for finding $N(x)$ and $h(x)$ from $\tildeC_1(x)$ and $\tildeC_2(x)$.
The function $N(x)\geq 0$ in \eqref{e49}
is the number of small particles per unit volume around point $x\in D$.
We want to prove that one can choose the functions $h_1(x)$ and $h_2(x)$
so that $N(x)$ is positive, $\tildeC_2\leq 0$,  $\tildeC_1, \tildeC_2$
are given functions, and  \eqref{e47} holds.
  
Let
\be\label{e50}
  \frac{|S|}{C} :=b>0, \qquad H:=bh(x)=H_1+iH_2. \ee
Then the first equation \eqref{e49} implies
\be\label{e51}
  CN(x)=\frac{\tildeC_1+i\tildeC_2}{H}+\tildeC_1+i\tildeC_2. \ee
Since $CN(x)>0$, and $N(x)>0$, the real part of the right side of 
\eqref{e51} should
be positive:
\be\label{e52}
  \frac{\tildeC_1 H_1+\tildeC_2H_2}{H^2_1 + H^2_2} +\tildeC_1>0, \ee
and the imaginary part should vanish:
\be\label{e53}
  \frac{\tildeC_2H_1-\tildeC_1H_2}{H^2_1+H^2_2}+\tildeC_2=0.\ee
Condition \eqref{e53} holds if 
\be\label{e54}
  \tildeC_1=\tildeC_2\frac{H^2_1+H^2_2+H_1}{H_2}, \qquad H_2\not=0.\ee
Using \eqref{e54}, write \eqref{e52} as
\be\label{e55}
  \frac{\tildeC_2}{H_2}\cdot[\frac { (H^2_1+H^2_2+H_1)^2+H_2^2}{ 
H^2_1+H^2_2}]>0.\ee
It follows from \eqref{e55} that $H_2$ and $\tildeC_2$ are of the same 
sign. We may be interested in the materials for which the solution 
of the scattering problem is unique. Therefore we wish to satisfy 
conditions \eqref{e48}. The argument below shows that if  $Im\,q_0=0$ and 
$Im\,q<0$, then $\tildeC_2<0$.

Indeed,
if $Im\,q_0=0$ and $Im\,q<0$, then $H_2< 0$, and \eqref{e55} 
implies $\tildeC_2<0$. Conversely, if $\tildeC_2<0$ and $Im\,q_0=0$,
then \eqref{e55} implies $H_2< 0$. 
We conclude that if  $H_2<0$, and $H_1,H_2$ satisfy \eqref{e54},
where $\tildeC_2<0$ and $\tildeC_1$ are given, then the 
number $N(x)$ of
small particles per unit volume around point $x\in D$, calculated
by formula \eqref{e51},  is nonnegative (it can vanish around some 
points $x$), so that it has physical meaning.

If one embeds small particles of the same shape
with the density of their numbers $N(x)$ and chooses the boundary 
impedance $h(x)$ of these
particles so that the function $bh(x):=H(x):=H_1+iH_2$ satisfies the
condition $H_2(x)<0$, and if  equation \eqref{e51} holds, then the
material one obtains  by the embedding of these small particles into
$D$ will have the desired refraction coefficient $n(x)$, and the relative
total volume of these particles will be negligible.

\section*{Appendix}

\nd\textit{1. Uniqueness of the solution to the problem \eqref{e3}, 
\eqref{e4}, \eqref{e6}.}

If $u_1$ and $u_2$ solve \eqref{e3}, \eqref{e4}, \eqref{e6}, then 
$v:=u_1-u_2$ solves \eqref{e6} and staifies the radiation condition:
\bee
  \frac{\partial v}{\partial r}=ikv+o\left(\frac{1}{r}\right).
  \eqno{\hbox{(A.1)}}\eee
Multiply \eqref{e6}, where $v$ replaces $u$, by $\barv$, the overbar 
stands for complex conjugate, take complex conjugate of \eqref{e6} with 
$v$ replacing $u$, multiply it by $v$, subtract from the first equation, 
use Green's formula, and get
\bee
  \int_{|x|=r} (\barv v_r-v\barv_r)ds
 -\int_D(q_0-\barq_0)|v|^2dx=0.
  \eqno{\hbox{(A.2)}}\eee
Use (A.1) and let $r\to\infty$ in (A.2). This yields
\bee
  -2i\,Im\int_D q_0(x)|v|^2dx+2i\,k\int_{S^2} \bigg|A(\beta)\bigg|^2d\beta =0,
  \eqno{\hbox{(A.3)}}\eee
where
\bee
  v(\beta r)=A(\beta)\frac{e^{ikr}}{r}+o\left(\frac{1}{r}\right), 
  \qquad r:=|x|\to\infty.
  \eqno{\hbox{(A.4)}}\eee
\textit{If $Im\,q_0\leq 0$, then {\rm(A.3)} implies $A(\beta)=0$.}

\nd This, in turn, implies that $v$ satisfies the following relations: 
\bee
  (\nabla^2+k^2)v=0\hbox{\quad in\quad}D':=\R^3\setminus D,
  \qquad \lim_{r\to\infty}\int_{|x|=r}|v|^2ds=0.
  \eqno{\hbox{(A.5)}}\eee
From (A.5) one concludes $v=0$ in $D'$ (see, e.g., Lemma 1 
in \cite[p.25]{R190}). 
This and the unique continuation property for the solution of the 
homogeneous elliptic equation \eqref{e6} imply that $v=0$ in $\R^3$.
The proof is complete.

\nd\textit{2. Uniqueness of the solution to problem 
\eqref{e9}-\eqref{e11}.}

We argue as above and get
\bee
  -2i\,Im\int_Dq_0(x)|v|^2dx + 2ik\int_{S^2}|A(\beta)|^2d\beta
  -2i \sum^M_{m=1}\int_{S_m}Im\,h_m|v|^2ds=0, \qquad k>0.
  \eqno{\hbox{(A.6)}}\eee
If $Im\,h_m \leq 0$, and $Im\,q_0(x)\leq 0$, then (A.6) implies 
$A(\beta)=0$, $v\mid_{S_m}=0$, and this implies $v=0$ in 
$\R^3\backslash\bigcup^M_{m=1} D_m$. This implies $v=0$ in $\R^3$
by the argument, given at the end of n.1 of this Appendix.

\nd \textit{3. Estimates of $G(x,y)$ as $|x-y|\to0$ and as 
$|x-y|\to\infty$.}

We start with the usual integral equation for $G$:
 \bee
  G(x,y)=g(x,y)-\int_D g(x,t)q_0(t)G(t,y)dt,
  \qquad g(x,y):=\frac{e^{ik|x-y|}}{4\pi|x-y|}.
  \eqno{\hbox{(A.7)}}\eee
Equation (A.7) implies
 \bee\label{A.8}
  G(x,y)=g(x,y)[1+O(|x-y|)], \quad \hbox{\,\, as \,\,} \qquad |x-y|\to 0,
  \eqno{\hbox{(A.8)}}\eee
provided that
 \bee
 \sup_{x,y\in\R^3} \int_D \ \frac{|q_0(t)|dt}{|x-t||t-y|}\leq c,
  \eqno{\hbox{(A.9)}}\eee
where $c>0$ stands for various constants. We assume that (A.9) holds. 
For example, (A.9) holds if $q_0$ is bounded. 
We also have
 \bee
  \nabla_yG(x,y)=\nabla_y g(x,y)-\int_D g(x,t)q_0(t)\nabla_y G(t,y)dt.
  \eqno{\hbox{(A.10)}}\eee
This implies
 \bee
  \nabla_y G(x,y)=\nabla_y g(x,y)
  \left[ 1+O\left(|x-y|^2\ln\frac{1}{|x-y|}\right)\right]
  \qquad \hbox{\quad as\quad} |x-y|\to 0.
  \eqno{\hbox{(A.11)}}\eee
Indeed,
 \bee
  \nabla_y g(x,y)=ikg(x,y) \left(1-\frac{1}{ik|x-y|}\right)
  \frac{y-x}{|y-x|},
  \eqno{\hbox{(A.12)}}\eee
so
 \bee
  |\nabla_y g(x,y)|=O(|x-y|^{-2}), \qquad \hbox{\quad as\quad} |x-y|\to 0,
  \eqno{\hbox{(A.13)}}\eee
and
 \bee
  \bigg|\int_D g(x,t)q_0(t) \nabla_y G(t,y)dt\bigg|
  \leq c \int_D \frac{dt}{|x-t||t-y|^2} \leq c\ln\frac{1}{|x-y|}
   \qquad \hbox{\quad as\quad} |x-y|\to 0.
  \eqno{\hbox{(A.14)}}\eee
Note that
 \bee
  g(x,y)=g_0(x,y) [1+O(ka)],\qquad g_0(x,y):=\frac{1}{4\pi|x-y|},
  \qquad \hbox{\quad if\quad} |x-y|<a.
  \eqno{\hbox{(A.15)}}\eee
Now, let $|x-y|\to\infty$. Assume that $y$ is in a bounded domain. Then 
(A.7) implies that
 \bee
  |G(x,y)|+|\nabla_y G(x,y)|=O\left(\frac{1}{|x|}\right),
  \qquad \hbox{\quad if\quad} |x|\to\infty, \qquad |y|\leq c.
  \eqno{\hbox{(A.16)}}\eee


\end{document}